\documentclass[prc,preprint,showpacs,aps,superscriptaddress,floatfix]{revtex4}

\usepackage[dvips]{color}
\usepackage{graphicx}

\begin{document}

\title{Alpha-decay lifetimes semiempirical relationship \\including shell
effects}

\author{D. N. Poenaru}
\email[]{Dorin.Poenaru@nipne.ro}

\affiliation{
Horia Hulubei National Institute of Physics and Nuclear
Engineering (IFIN-HH), \\RO-077125 Bucharest-Magurele, Romania}

\author{R. A. Gherghescu}
\affiliation{
Horia Hulubei National Institute of Physics and Nuclear
Engineering  (IFIN-HH), \\RO-077125 Bucharest-Magurele, Romania}

\author{N. Carjan}

\affiliation{Centre d'Etudes Nucl\'eaires de Bordeaux Gradignan  (CENBG), BP
120,
F-33175 Gradignan Cedex, France}


\begin{abstract}
A new version of the semiempirical formula based on fission approach of
alpha decay is derived, by using the optimum values of the fitting
parameters determined for even-even nuclei, combined with hindrance factors
for even-odd, odd-even, and odd-odd nuclides.
The deviations from experimental data for two regions of nuclear chart (493
alpha emitters with $Z=52-118$ and 142 transuranium nuclei including
superheavies ($Z=92-118$), respectively) are
compared with those obtained by using the universal curve and the
Viola-Seaborg semiempirical relationship.
\end{abstract}


\maketitle


The existence of a fission barrier produced by shell effects in the region
of superheavy nuclei \cite{gre96ijmp,mos69zp} continues to be proved by the
successful synthesis of the heaviest nuclides
\cite{hof00rmp,oga05pr,mor04jpsj}, which stimulate a corresponding
theoretical development (see
\cite{moh06pr,xu06pr,ghe06pr,ghe05pr} and the references therein).
In 1969 it was shown that fission barriers as
high as 8 to 12 MeV develop due to the shell corrections around
double shell closures which appeared in the standard shell model at 
$Z=114$, $N=184$. Nowadays some other magic numbers have been proposed, e.g.
$Z=126$.
The majority of these proton-rich nuclei are mainly decaying by
$\alpha$-particle emission, motivating an important effort to derive
different methods allowing to estimate the half-lives \cite{roy00jpg,par05app}
and to perform systematic studies in this region
e.g. \cite{gup05nds,gam05pr,cho06pr}.

Three such methods based on fission theory extended to a very large asymmetry
(the analytical superasymmetric fission (ASAF) model
\cite{p195b96b,p160adnd91}, the universal (UNIV) curve \cite{p167ps91}, 
and the semiempirical relationship (SemFIS) \cite{p83jpl80,p98cpc82})
have been compared recently \cite{p269jpg06,p268pr06}. The kinetic energy
of the emitted particle, $E_\alpha = QA_1/A$, is given at the input.
It was shown that the description of data in the vicinity of the
magic proton and neutron numbers, where the errors of the other
relationships are large, was improved by introducing the SemFIS formula, 
in which the action integral is multiplied by a
second order polynomial, $\chi=\chi(x,y)$, 
in the variables $x$ and $y$ expressing the
``distance'' from the nearest magic plus one neighbours
$x=(N-N_i)/(N_{i+1} - N_i)$ and $y=(Z-Z_i)/(Z_{i+1} - Z_i)$, where $N_i > N
> N_{i+1}$ and $Z_i > Z > Z_{i+1}$.
In this way not only the $Z$ dependence (present in the majority of
semiempirical relationships) but also the strong
influence of the neutron shell effects are taken into account. 
A computer program \cite{p98cpc82}
allows to change automatically the fit parameters, every time a
better set of experimental data is available. This polynomial approximation
needs six parameters in each group of even-even, even-odd, odd-even and
odd-odd alpha emitters. The purpose of the present work is to derive an
alternative with less number of parameters and to compare the results with
those obtained by using the universal curve.
The same six quantities determined for even-even nuclei should be used
in the other groups, only completed in the expression of $\log T$ 
by a different hindrance factor.

We study a spontaneous process of $\alpha$-decay in which
a {\em parent} nucleus $A,Z$ in its ground state is split into two fragments
$A_1=A-4, Z_1=Z-2$ and $A_2=4, Z_2=2$ 
\begin{equation} 
AZ \rightarrow \; \; ^{A_1}Z_1 + \; ^{A_2}Z_2 
\end{equation}
in a way that conserves the hadron numbers $A=A_1+A_2$, $Z=Z_1+Z_2$. The
heavy fragment $A_1,Z_1$ is called {\em daughter} and the light one {\em
emitted $\alpha$-particle}. The partial decay half-life $T$
of the parent nucleus is related to the
disintegration constant $\lambda $ of the
exponential decay law in time by the relationship
\begin{equation}
\frac {\ln 2}{T} = \lambda  
 \label{half}
\end{equation}
The quantum mechanical tunnelling process
leads to a relatively simple expression
for $\lambda$ as the product of the three model dependent quantities 
\begin{equation}
\lambda = \nu S P
 \label{la}
\end{equation}
where $\nu $ is the frequency of assaults, $S$ is the preformation
probability of the $\alpha$-particle at the nuclear surface, and $P$ is the
penetrability of the external part of the barrier \cite{p167ps91,p195b96b}.

Very frequently the penetrability is calculated by using the one-dimensional 
Wentzel--Kra\-mers--Brillouin (WKB) approximation
\begin{equation}
P = \exp (-K) ; \; \; K=\frac{2}{\hbar}\int\limits_{R_a}^{R_b}
\sqrt{2B(R)E(R)} dR
 \label{pen}
\end{equation}
where $B$ is the nuclear inertia, approximated by the reduced 
mass $\mu=mA_1A_2/A$, $m$ is the nucleon mass, $E$ is the potential
energy from which the $Q$-value has been subtracted out, $R_a$ and $R_b$ are
the classical turning points, and $K$ is the action integral.

The equations \ref{half}-\ref{pen} are used as a 
starting point by the three methods mentioned above (ASAF, UNIV, and
SemFIS) based on fission theory of $\alpha$-decay. The zero point vibration
energy is $E_v=h\nu/2$ in which $h$ is the Plank constant. The outer potential
barrier is of Coulomb nature, $E(R)=e^2Z_1Z_2/R -Q$, with the
electron charge $e=1.43998$~MeV$\cdot$fm. The other 
numerical constants are given by $(1/2)h \ln 2$ = 1.4333$\cdot
10^{-21}$ MeV$\cdot$s and $2\sqrt{2m} /\hbar$ = 0.43921
MeV$^{-1/2}$fm$^{-1/2}$.
  
It is instructive to outline the derivation of the UNIV formula.
For $\alpha$-decay of even-even nuclei the eqs.~\ref{half}-\ref{pen} 
allow us to obtain a simple relationship of 
the decimal logarithm of the half-life
\begin{equation}
\log T = -\log P + c_{ee}
\label{lgt}
\end{equation}
where the additive constant
$c_{ee}=\log S_{\alpha} - \log \nu + \log (\ln 2) = -20.325$ if we
are making two approximations: $S_{\alpha}=0.0180302$ and
$\nu=10^{22.01}$~s$^{-1}$. For even-odd, odd-even and odd-odd nuclei
we replace $c_{ee}$ by $c_{eo} = c_{ee} + h_{eo}$, $c_{oe} = c_{ee} +
h_{oe}$, and $c_{oo} = c_{ee} + h_{oo}$, respectively, where $h_{eo},
h_{oe}, h_{oo}$ are the mean values of the hindrance factors in these groups
of nuclides. In a doubly logarithmic scale the equation~\ref{lgt} represents
a straight line with a slope equal to unity. 
The penetrability of an external Coulomb barrier, having as the first turning
point the separation distance at the touching configuration $R_a=R_t=R_1+R_2$ 
and the second one defined by $e^2Z_1Z_2/R_t=Q$,
may be found analytically as
\begin{equation}
-\log P = 0.22873({\mu}_{A}Z_{1}Z_{2}R_{b})^{1/2} \left [
\arccos \sqrt{r} - \sqrt{r(1-r)}\right ]
\end{equation}
where $r=R_{t}/R_{b}$, $R_{t}=1.2249(A_{1}^{1/3}+A_{2}^{1/3})$, and
$R_{b}=1.43998Z_{1}Z_{2}/Q$. We use the liquid drop model radius constant
$r_0=1.2249$~fm.

A great number of alpha emitters are known 
\cite{roe96mb,ryt91adnd,aud03np,aud03np1,gup05nds,gam05pr} 
as it may be seen in
figure~\ref{texp} where we plotted the experimental half-lives of 162
even-even, 122 even-odd, 115 odd-even, and of 94 odd-odd nuclides versus the
neutron number of the daughter nucleus. The very strong neutron shell effect
is clearly seen at the magic daughter number $N_d=126$, where the shortest
half-lives have been measured.

The corresponding universal curves for $\alpha$-decay of the same nuclides
as in figure~\ref{texp} in four groups of even-even, even-odd, odd-even and
odd-odd nuclei are displayed in figure~\ref{unia}. The following hindrance
factors have been used:  $h_{oe}=0.445$, $h_{e0}=0.294$, $h_{oo}=0.842$,
for even-odd, odd-even, and odd-odd nuclei, respectively.
Qualitatively one can see that the data for even-even nuclei are well
described by the universal curve. We can evaluate the standard
root-mean-square (rms) deviation of $\log T$ values:
\begin{equation}
\sigma = \left\{\sum_{i=1}^n[\log
(T_i/T_{exp})]^2/(n-1)\right\}^{1/2}
\end{equation}
leading to  $\sigma = 0.352, 0.612, 0.546, 0.841$ for the 162 even-even, 122
even-odd, 115 odd-even, and of 94 odd-odd nuclides, respectively.

Despite the fact that a main part of the strong shell effect present in the
$Q$-value was accounted for, there are still some systematic errors in the
vicinity of the magic number of neutrons as may be seen in the 
plot of $\log T$ --
$\log T_{exp}$ against the neutron number of the daughter (see
 figure~14 of
the ref. \cite{p269jpg06}). 

The same kind of underestimation around neutron magic numbers is also
present in the semiempirical formulae.
A typical example is the
Viola-Seaborg \cite{vio66jinc} semiempirical relationship 
\begin{equation}
\log T = (aZ -  b)Q^{-1/2} - (cZ + d)
\end{equation}
We shall
use for $Z \leq 82, N\leq 126$ the parameter values $a=2.42151$;
$b=62.3848$; $c=0.59015$, $d=4.2109$, and for $Z > 82, N > 126$ the
recently published \cite{par05app} ones $a=1.3892$; $b=-13.862$;
$c=0.1086$, $d=41.458$.
It gives excellent agreement
in the region of actinides but it underestimates the lifetimes of
lighter nuclei.
It is very frequently used, particularly in the region of superheavy
nuclei.

The behaviour around magic numbers can be improved by using the
SemFIS formula in which we explicitly take into account the shell effect by
multiplying the action integral, $K$, with the polynomial $\chi=\chi(x,y)$
mentioned above:
\begin{equation}
\log T = 0.43429\chi (x,y) \cdot K - 20.446 + H^f
\label{eq:pf}
\end{equation}
where $H^f$ is a hindrance factor which takes different values 
$H^f_{ee}=-0.025$ for even-even emitters, $H^f_{eo}=0.420$ for e-o, 
$H^f_{oe}=0.280$ for o-e, and $H^f_{oo}=0.810$ for o-o ones.
\begin{eqnarray}
K & = & 2.52956 Z_{1}[A_{1}/(AQ)]^{1/2}[\arccos \sqrt{r} 
- \sqrt{r(1-r)}] \; ; \nonumber \\
   & r & =0.423Q(1.5874+A_{1}^{1/3})/Z_{1} 
\end{eqnarray}
The numerical coefficient $\chi$, close to unity, is a
second-order polynomial
\begin{equation}
\chi=B_1+ x(B_2+ xB_4) + y(B_3+ yB_6) + xyB_5
\end{equation}
with coefficients obtained by fitting the data plotted in figure~\ref{texp} 
for 163 even-even emitters: $B_1=0.987389$, $B_2=0.009520$, 
$B_4=0.033711$, $B_3=0.035249$,  $B_6=-0.038941$, $B_5=0.019451$.
The reduced variables are defined 
\begin{equation}
x \equiv (N-N_i)/(N_{i+1} - N_i) \; ; \; N_i < N \le N_{i+1}
\end{equation}
\begin{equation}
y \equiv (Z-Z_i)/(Z_{i+1} - Z_i) \; ; \; Z_i < Z \le Z_{i+1}
\label{eq:pl}
\end{equation}
with $N_i=...., 51, 83, 127, 185, 229, .....$ , 
$Z_i=...., 29, 51, 83, 115, .....$
hence for the region of superheavy nuclei $x=(Z-83)/(127-83), \; \;
y=(N-127)/(185-127)$.

\begin{table}[th]
\caption{Standard deviations for semiempirical formula, and universal
curves in the region of all alpha emitters ($Z=52-118$).}

{\begin{tabular}{ll|r|r} \toprule
 n &parity &  $\sigma_{univ}$ & $\sigma_{semFIS}$ \\  \hline
 162~~ &e-e&   0.352&   0.233     \\  \hline 
 122~~ &e-o&   0.612&   0.515   \\ \hline 
 115~~ &o-e&   0.546&   0.497   \\ \hline 
  ~~94~~ &o-o&   0.841&   0.709  \\  \botrule
\end{tabular}}
\label{tab1}
\end{table}
The improvement can be seen by comparing the standard deviations of UNIV and
SemFIS for the large set of data in table~\ref{tab1}. The figure~\ref{sem2}
shows that there are nolonger large systematical discrepancies around magic
numbers, because shell effects were taken into account by SemFIS. The
example are given for the best results obtained in the region of even-even
(top) and odd-even (bottom) nuclei.

In order to illustrate the possibility of using a different set of parameter
values, we apply the semFIS formula for transuranium nuclei (the set of 47
even-even ($Z=92, 118$), 45 even-odd ($Z=92, 114$), 25 odd-even
($Z=93, 115$), 25 odd-odd ($Z=93, 113$) nuclides). Now we obtain the 
following parameters: $B_1=0.985415$, $B_2=0.102199$, 
$B_4=-0.832081$, $B_3=-0.024863$,  $B_6=-0.681221$, $B_5=1.50572$
and $H^f_{ee}=0$, $H^f_{eo}=0.63$, $H^f_{oe}=0.51$, $H^f_{oo}=1.26$. 

The optimum hindrance factors for universal curves are 
$h_{ee}=0.100$, $h_{eo}=0.792$, $h_{oe}=0.642$, and $h_{oo}=1.512$. 

For Viola-Seaborg besides $a=1.3892$, $b=-13.862$, $c=0.1086$,
$d=41.458$ we introduce the additive hindrance
constants CV of $-0.073$ for e-e, $0.484$ for e-o,
$0.478$ for o-e, and $1.078$ for o-o nuclei.

\begin{table}[th]
\caption{Standard deviations for semiempirical formula, and universal
curves in the region of transuranium alpha emitters including 
superheavies ($Z=92-118$).}

\vspace*{3mm}
{\begin{tabular}{ll|r|r|r} \toprule
 n &parity & $\sigma_{VS}$ & $\sigma_{univ}$ &$\sigma_{semFIS}$ \\  \hline
 47~~ &e-e&0.282&0.267&0.164     \\  \hline 
 45~~ &e-o&0.562&0.554&0.540   \\ \hline 
 25~~ &o-e&0.608&0.543&0.514   \\ \hline 
 25~~ &o-o&0.597&0.456&0.492  \\  \botrule
\end{tabular}}
\label{tab2}
\end{table}

From the table~\ref{tab2} one can see how the universal curves reproduce
better than the Viola-Seaborg formula the experimental data. A similar
property is exhibited by the semFIS formula, which behaves very well for
even-even alpha emitters. Moreover, the dependence on the proton and neutron
magic numbers of the semiempirical formula may be exploited to obtain
informations about the values of the magic numbers which are not well known
until now.

In conclusion the new version of the SemFIS formula, having a smaller number
of fitting parameters, gives slightly better results than the universal
curve, both for a large number (493) of alpha emitters in the whole nuclear
chart ($Z=52-118$) and for 142 transuranium nuclei including superheavies
($Z=92-118$).




\bigskip \bigskip

\section*{FIGURE LEGEND }

\bigskip

\noindent {\bf Figure 1} ~~~~~
Experimental half-lives for $\alpha$-decay of 162 even-even (E-E),
122 even-odd (E-O), 115 odd-even, and of 94 odd-odd (O-O) nuclides
versus the neutron number of the daughter nucleus. 
The vertical bars correspond to spherical and deformed neutron 
magic numbers of the daughter nuclei $N_d=50, 82, 126, 152, 162, 172$.
They span a wide range of values between $10^{-7}$ and $10^{25}$~s. 
The particularly strong shell effect at $N_d=126$ is very clearly seen.

\bigskip 

\begin{figure}[bh]
\centerline{\includegraphics[width=14cm]{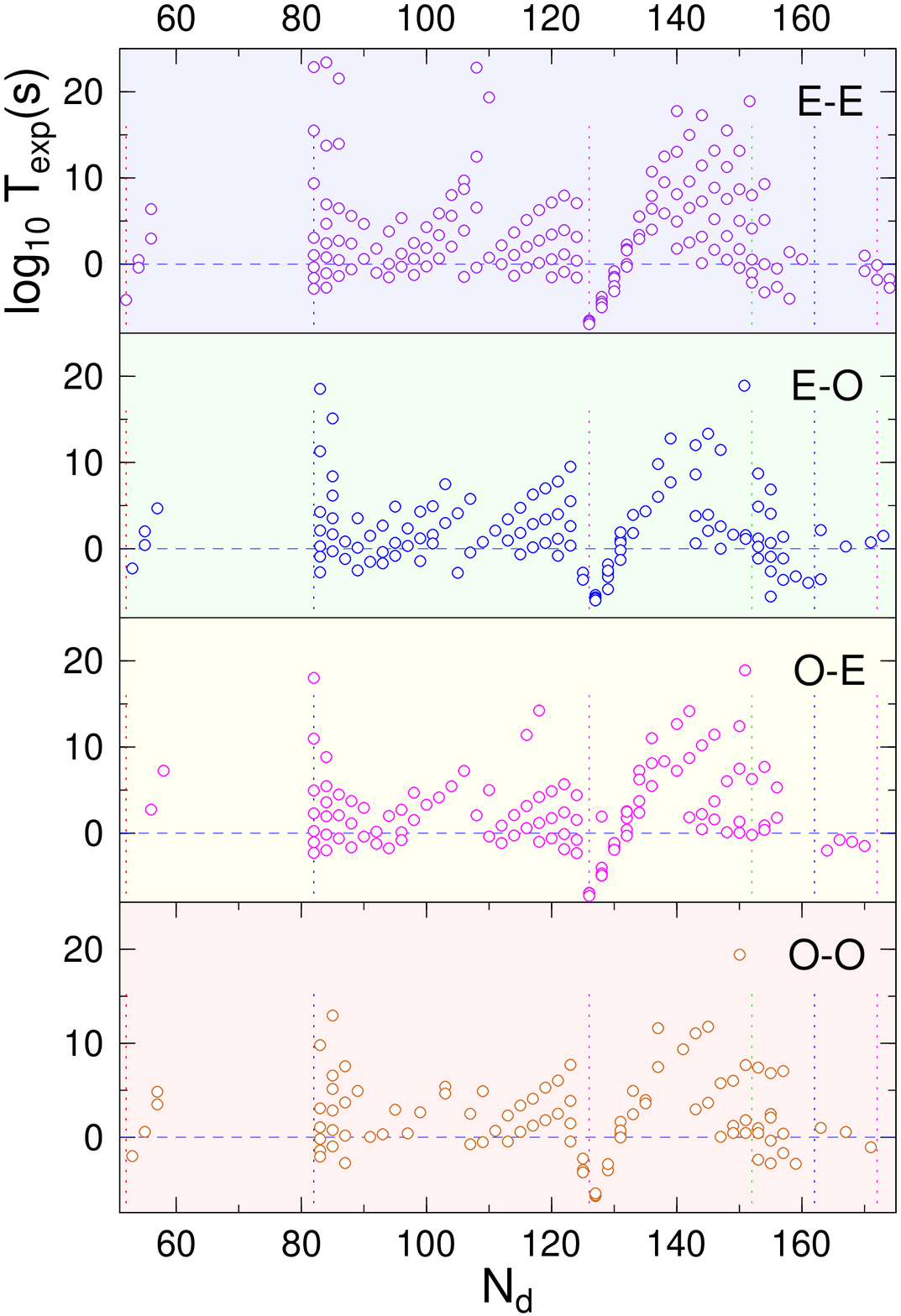}}
\caption{
\label{texp}}
\end{figure}

\begin{figure}[bh]
\vspace*{2cm}
\centerline{\includegraphics[width=16cm]{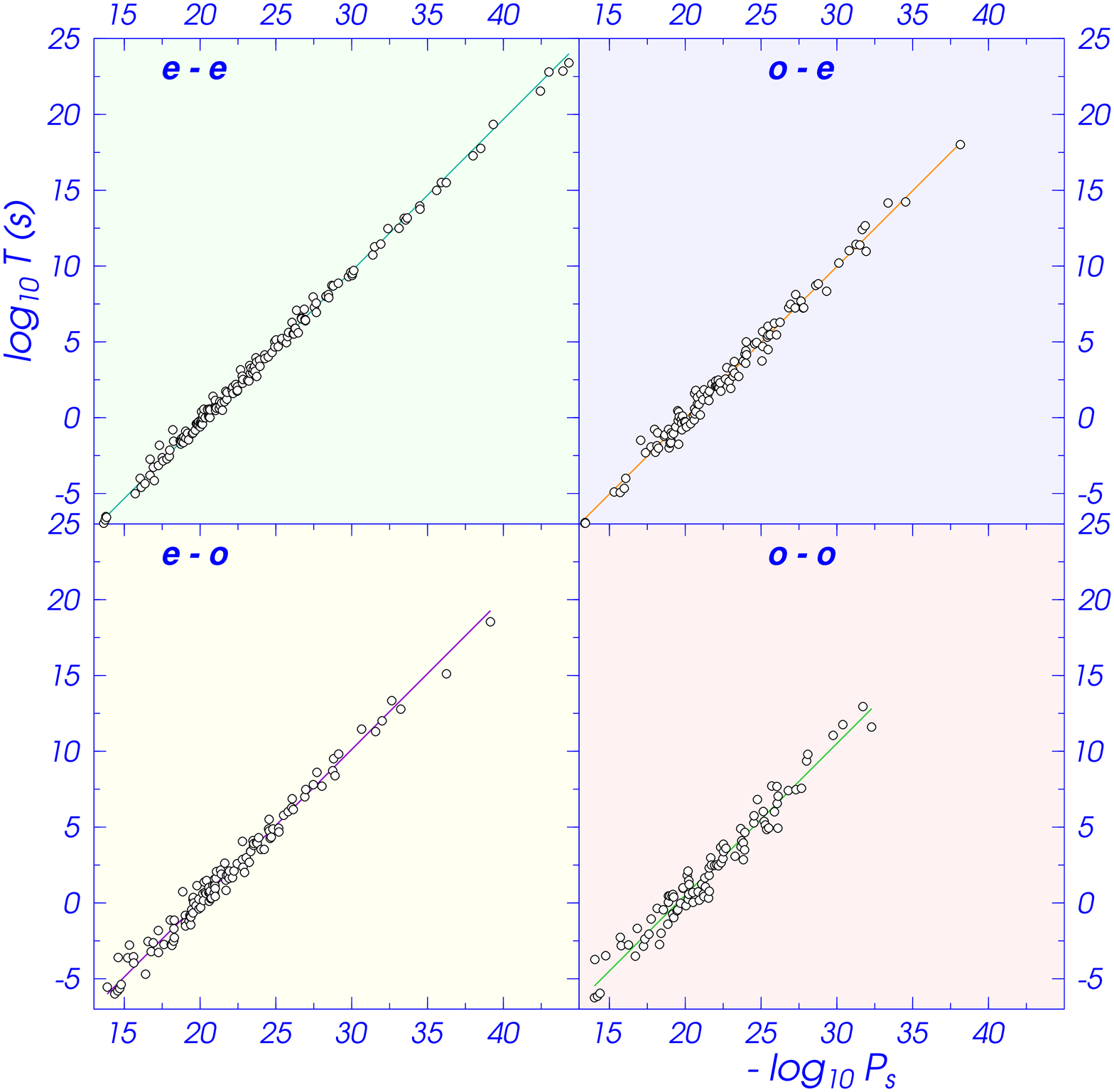}}
\caption{
\label{unia}}
\end{figure}

\begin{figure}[th]
\centerline{\includegraphics[width=14cm]{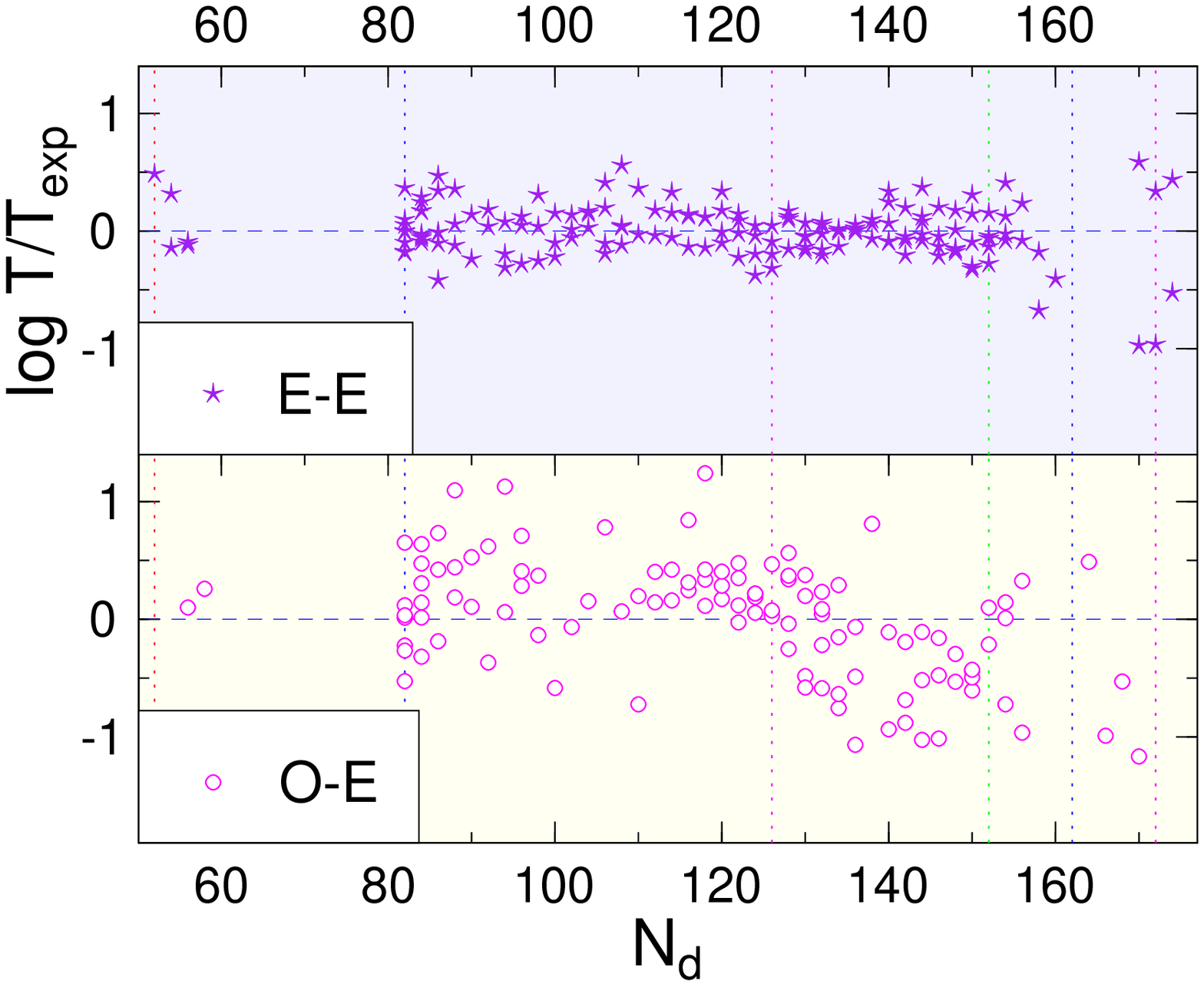}}
\caption{
\label{sem2}}
\end{figure}

\noindent {\bf Figure 2} ~~~~~
Universal curves for $\alpha$-decay of the same nuclides as in
figure~\ref{texp} 
in four groups of even-even, even-odd, odd-even and odd-odd nuclei.

\bigskip 

\noindent {\bf Figure 3}  ~~~~~ 
The deviations of $\alpha$-decay half-lives calculated with the SemFIS
semiempirical formula from the experimental values  for even-even (top) and
odd-even (bottom) nuclei.
The vertical bars correspond to spherical and deformed neutron
magic numbers of the daughter nuclei $N_d=50, 82, 126, 152, 162, 172$.
There are nolonger large systematical discrepancies around magic numbers, 
because shell effects were taken into account by SemFIS.
Calculations are performed with the new constants adjusted to fit the data
of even-even nuclei.

\newpage

\bigskip \bigskip

\newpage

\end{document}